\documentclass[sigconf]{acmart}

\usepackage{graphicx}
\usepackage{slashbox}
\usepackage{pict2e}
\usepackage[utf8]{inputenc}
\usepackage{url}
\usepackage{amsmath}
\usepackage[T1]{fontenc}
\usepackage{multirow}
\usepackage{adjustbox}
\usepackage{lipsum,multicol}
\usepackage{amssymb}
\usepackage{pifont}
\newcommand{\cmark}{\ding{51}}%
\usepackage{mathtools,etoolbox}
\DeclarePairedDelimiterX{\abs}[1]{\lvert}{\rvert}{\ifblank{#1}{{}\cdot{}}{#1}}

\usepackage{subcaption}
\usepackage{mwe}

\newcommand{\reffig}[1]{Fig.~\ref{#1}}
\newcommand{\reftab}[1]{Table~\ref{#1}}
\newcommand{\refsec}[1]{Section~\ref{#1}}

\renewcommand{\theenumi}[1]{Appendix~\alph{enumi}}

\def\BibTeX{{\rm B\kern-.05em{\sc i\kern-.025em b}\kern-.08em
    T\kern-.1667em\lower.7ex\hbox{E}\kern-.125emX}}

\setcopyright{rightsretained}

\acmDOI{}

\acmISBN{XXX-X-XXXXX-XXX-X}

\acmConference[]{()}{2020}{} 
\acmYear{2020}

\begin{document}
\title{Knowledge Graphs for Processing Scientific Data:\\ Challenges and Prospects}
  

\author{Masoud Salehpour}
\affiliation{\institution{University of Sydney}}
\author{Joseph G. Davis}
\affiliation{\institution{University of Sydney}}

\begin{abstract}
There is growing interest in the use of Knowledge Graphs (KGs) for the representation, exchange, and reuse of scientific data. While KGs offer the prospect of improving the infrastructure for working with scalable and reusable scholarly data consistent with the FAIR (Findability, Accessibility, Interoperability, and Reusability) principles, the state-of-the-art Data Management Systems (DMSs) for processing large KGs leave somewhat to be desired. In this paper, we studied the performance of some of the major DMSs in the context of querying KGs with the goal of providing a finely-grained, comparative analysis of DMSs representing each of the four major DMS types. We experimented with four well-known scientific KGs, namely, Allie, Cellcycle, DrugBank, and LinkedSPL against Virtuoso, Blazegraph, RDF-3X, and MongoDB as the representative DMSs. Our results suggest that the DMSs display limitations in processing complex queries on the KG datasets. Depending on the query type, the performance differentials can be several orders of magnitude. Also, no single DMS appears to offer consistently superior performance. We present an analysis of the underlying issues and outline two integrated approaches and proposals for resolving the problem.

\end{abstract}

\maketitle

\section{Introduction}
\label{sec::introduction}
The massive increase in the \textit{accessibility} and \textit{heterogeneity} of scientific and statistical datasets have heightened interest in the use of \textit{Knowledge Graphs} (\textit{KGs}) for their representation and exchange. For instance, a range of scientific providers such as NCBI\footnote{\url{https://www.ncbi.nlm.nih.gov/}}, NLM\footnote{\url{http://id.nlm.nih.gov/mesh}}, Neurocommons, and Protein Data Bank Japan, to name a few, have made scientific KGs available for public access. Scientific KGs are \textit{directed}, \textit{edge-labeled} graphs created to accumulate, represent, and exchange scientific knowledge (usually domain-specific) in which nodes represent the real world entities of interest (e.g., genes, proteins, drugs, etc.) and edges represent \textit{interrelations} between these entities. Knowledge in these graphs is usually composed of simple statements, such as \mbox{``Acetaminophen~~is\_a~~drug''} in which ``Acetaminophen'' and ``drug'' are nodes and ``is\_a'' is a label of a directed edge. We refer to each statement as a triple. A KG may contain thousands to billions of \textit{triples} generally available in the form of \textit{RDF}\footnote{The World Wide Web Consortium (W3C) has recommended the Resource Description Framework (RDF) as a directed and labeled graph-like structure for \textit{representation}, \textit{integration}, and \textit{exchange} of the content of a KG using a large set of triples of the form $<$subject predicate object$>$.} datasets that can be queried using a standard RDF query language such as \textit{SPARQL}.\footnote{\url{http://www.w3.org/TR/rdf-sparql-query/}}

A scientific KG itself is not a goal, but when it becomes Findable, Accessible, Interoperable, and Reusable (aka, FAIR principles~\cite{fair2016}), it has the potential to be the key driving force behind further knowledge discovery and integration. Through this paper we hope to initiate a discussion on the efficacy and efficiency of services offered by the current generation of Data Management System (DMS) under conditions when scientific KGs are characterized by (i) large volume of data, (ii) a large and diverse number of queries (either ad-hoc or batch), (iii) separation of the accessibility from the internal/physical storage, (iv) concurrent access, and (v) heavily interrelated and constrained data. Among these, executing a large number of queries (especially at scale) is a critical requirement for scientific data processing. DMS designers have employed a variety of \textit{design choices and architectures} for querying KGs over the past few years. For example, several exhaustive indexing strategies~\cite{Thomasindex}, compression techniques, and dictionary encoding (to keep space requirements reasonable for excessive indexing) have been implemented by major DMSs such as multiple bitmap indexes of Virtuoso or dictionary-based lexical values encoding of Blazegraph. A range of research prototypes have also been presented. For instance,~\cite{tamer2019} proposed a workload-adaptive and self-tuning DMSs using physical clustering of the underlying data and~\cite{RDF3x} proposed the ``RISC-style'' architecture to leverage multiple query processing algorithms and optimization. However, the absence of an \textit{explicit schema} and the \textit{heterogeneity} of scientific KG content pose challenges to DMSs for \textit{querying} these KGs efficiently since DMSs typically cannot make any \mbox{a priori} assumption about the structure of the content~\cite{saleem,IBMapple}. The problem of querying large scientific KGs efficiently calls for greater research attention.

In this paper, we present experimental evidence to suggest that the current generation of DMS tools are limited in their ability to support scientists in their research using KGs. We provide a fine-grained, comparative performance analysis of the major DMS types in the context of processing scientific KGs. For our experiment, we selected Virtuoso, Blazegraph, RDF-3X, and MongoDB as representative DMSs. Virtuoso was selected since it is already employed as the DMS of choice for a broad range of scientific KGs (e.g., the Linked Data for the Life Sciences project\footnote{\url{https://bio2rdf.org/sparql}}). Blazegraph\footnote{It is alleged that Blazegraph acquihired by Amazon and the Amazon Neptune is based on Blazegraph.} was selected since it is the DMS behind Wikidata,\footnote{\url{https://query.wikidata.org/}} (a KG constructed from the content of Wikimedia sister projects including Wikipedia, Wikivoyage, Wiktionary, and Wikisource). RDF-3X was selected since it is one of the most optimized open-source prototypes which has been employed in many studies as a baseline such as~\cite{watdiv}. The efficacy of document-stores for executing queries against scientific KGs has not been researched extensively. However, some academic prototypes such as~\cite{Jignesh} employed document-stores in other similar contexts. MongoDB was selected as a representative document-store since it is considered to be the leader in this class of tools~\cite{Jignesh}.

We loaded four well-known scientific KGs, namely, Allie\footnote{\url{http://allie.dbcls.jp/}}, \mbox{Cellcycle}\footnote{\url{ftp://ftp.dbcls.jp/togordf/bmtoyama/cellcycle/}} (aka, Semantic Systems Biology-CCO), DrugBank\footnote{\url{https://download.bio2rdf.org/files/current/drugbank/drugbank.html}}, and \mbox{LinkedSPL}\footnote{\url{https://download.bio2rdf.org/files/current/linkedspl/linkedspl.html}} into the DMSs separately. Relevant SPARQL queries were executed over each of the DMSs and query execution times computed to analyze the performance of each DMS. Our contributions include:

\begin{itemize}
\item Comparative performance analysis and experimental evaluation of major DMSs in supporting scientific KG query processing
    \item Providing explanations for the observed strengths and limitations of the different DMSs
    \item Analyzing the underlying issues related to the performance differentials and proposing approaches to resolve the problem
\end{itemize}

The remainder of this paper is organized as follows. In~\refsec{sec::preliminaries}, we provide some preliminary information about KG query types.~\refsec{sec::Experimental-Setting} presents our experimental setup including the scientific KG characteristics, computational environment, DMSs configuration, indexing, and data loading process. In~\refsec{sec::Evaluation}, results of the query processing and related analyses are presented. We summarize the lessons learned from our research and outline two proposals to resolve the problem in \refsec{sec::discussion}.~\refsec{sec::related_work} highlights related work. We present our conclusions and future work in~\refsec{sec::conclusion}.

\begin{figure}[t]
\centering\includegraphics[width=0.5\textwidth]{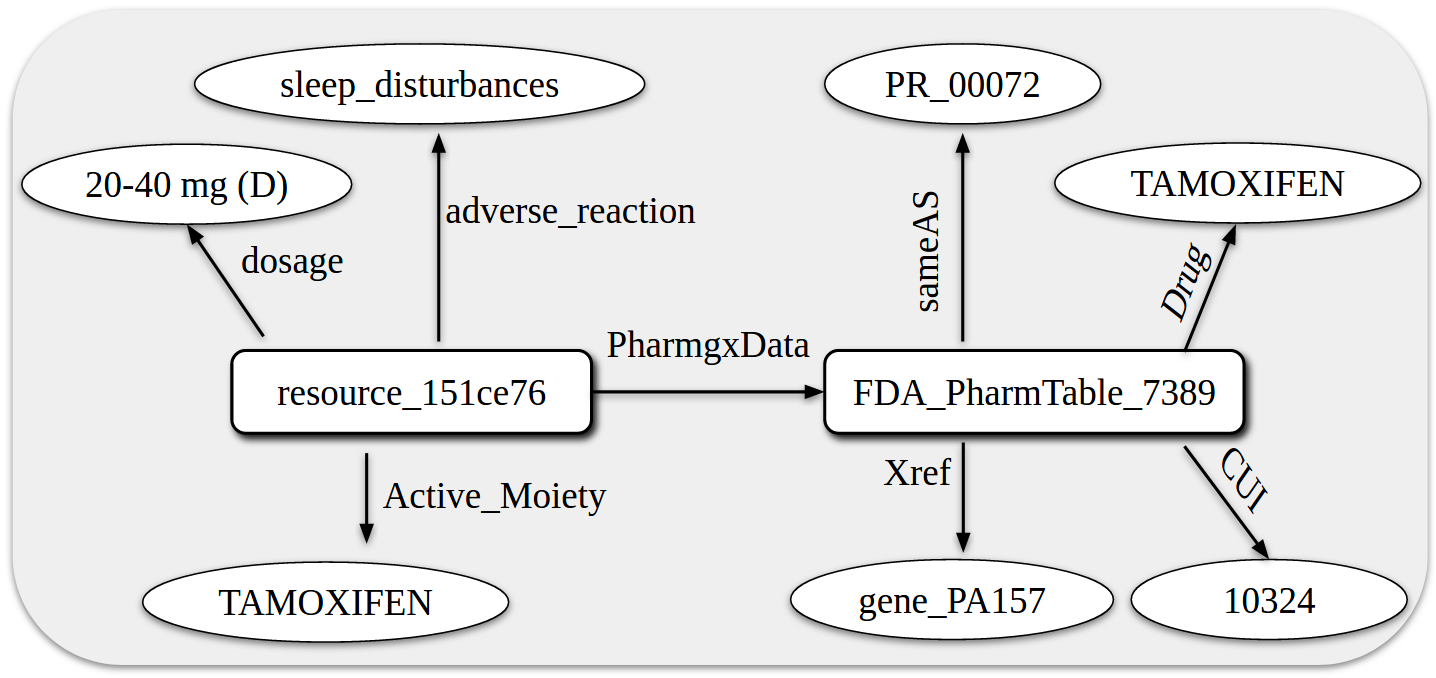}
\caption{An example of a simple scientific Knowledge Graph (based on the LikedSPL KG).}
\label{fig::kg}
\end{figure}

\section{Knowledge Graph Query Types}
\label{sec::preliminaries}
In this section, we present some preliminary information about the major KG query types using a human-readable example depicted in \reffig{fig::kg}. This is a small extract from the LinkedSPL KG which includes all sections of FDA-approved prescriptions and over-the-counter drug package inserts from DailyMed. The content of this KG subset can be represented by the following RDF triples:

\begin{verbatim}
resource_151ce76     dosage       20-40 mg (D)
resource_151ce76 adverse_reaction sleep_disturbance
resource_151ce76   Active_Moiety  TAMOXIFEN
resource_151ce76   PharmgxData    FDA_PharmTable_6540
FDA_PharmTable_6540  SameAs       PR_00072
FDA_PharmTable_6540  Drug         TAMOXIFEN
FDA_PharmTable_6540  CUI          10324
FDA_PharmTable_6540  Xref         gene_PA157
\end{verbatim}

An example of a query\footnote{We assume that the reader is familiar with the basic concepts of querying KG, e.g., the SELECT clauses.} is given below. It asks for the dosage of the subject ``resource\_151ce76''. ``?dosage'' is a variable to return the associated value as the result (i.e., ``20-40 mg (D)''). Queries may contain a set of triple patterns such as \textit{\mbox{``resource\_151ce76 dosage ?dosage''}} in which the subject, predicate, and/or object can be a variable.

\begin{verbatim}
    SELECT ?dosage
    WHERE {
    resource_151ce76 dosage ?dosage . 
    }
\end{verbatim}

Each triple pattern typically returns a subgraph. This resultant subgraph can be further \textit{joined} with the results of other triple patterns to return the final resultset. In practice, there are three major types of join queries: (i) subject-subject joins (aka, star-like), (ii) subject-object joins (aka, chain-like or a path), and (iii) tree-like (i.e., a combination of subject-subject and subject-object joins).

\textbf{Subject-subject joins.} A subject-subject join is performed by a DMS when a KG query has at least two triple patterns such that the predicate and object of each triple pattern is a given value (or a variable), but the subjects of both triple patterns are replaced by the \textit{same} variable. For example, the following query looks for all subjects for which their dosage and adverse reactions are equal to the given values (the result will be ``resource\_151ce76'').

\begin{verbatim}
    SELECT ?x
    WHERE {
    ?x  dosage            "20-40 mg (D)"      .
    ?x  adverse_reaction  "sleep_disturbance" .
    }
\end{verbatim}

\textbf{Subject-object joins.} A subject-object join is performed by a DMS when a KG query has at least two triple patterns such that the subject of one of the triple patterns and the object of the other triple pattern are replaced by the same variable. For example, the following query looks for all subjects that are connected to the FDA's pharmacogenomic biomarker table through ``PharmgxData'' predicate and their CUI\footnote{CUI (aka, RxCUI) is a unique, unambiguous identifier that is assigned to an individual drug} is equal to ``10324'' (``resource\_151ce76'' is the result).

\begin{verbatim}
    SELECT ?y
    WHERE {
    ?x  CUI          10324  .
    ?y  PharmgxData  ?x     .
    }
\end{verbatim}

\textbf{Tree-like joins.} A tree-like join consists of a \textit{combination} of subject-subject and subject-object joins. For example, the following query looks for the ``Xref'' of all subjects that are connected to the FDA's pharmacogenomic biomarker table through ``PharmgxData'' predicate and have ``TAMOXIFEN'' as ``Active\_Moiety'' and also have ``CUI'' value of 10324 (the result will be ``gene\_PA157'').

\begin{verbatim}
    SELECT ?y
    WHERE {
    ?x  Active_Moiety  "TAMOXIFEN" .
    ?x  PharmgxData    ?z          .
    ?z  CUI            10324       .
    ?z  Xref           ?y          .
    }
\end{verbatim}

In addition to the query types, we provide a brief explanation of query selectivity and optional patterns. Each KG query contains a set of triple patterns in the form of \mbox{\textit{``subject~~~predicate~~~object''}}. The \textit{subject}, \textit{predicate}, and the \textit{object} part of a triple pattern maybe concrete (i.e. bound) or variable (i.e. unbound). Sets of triple patterns specify the complexity of access to the underlying data. When the number of stored triples satisfying sets of triple pattern conditions is large as compared to the total number of stored triples, the corresponding query is considered to be low-selective~\cite{selectivity}. In other words, each query type can also be either high-selective or low-selective depending on the number of stored triples satisfying its triple pattern conditions. As explained previously, queries return resultsets only when the entire query pattern matches the content of the KG. However, some queries may contain optional patterns to allow KG queries to return a resultset even if the optional part of the query is not matched since completeness and adherence of KG content to their formal ontology specification is not always enforced.

\section{Experimental Setup}
\label{sec::Experimental-Setting}
\noindent In this section, we describe the scientific KGs. As well, our \textit{computational environment} and the DMS \textit{configurations} are described in detail.

\begin{table}
\centering
\begin{tabular}{l r r r r}
\hline
\backslashbox{KG}{Statistics} & Sub. (\#) & Pre. (\#)& Obj. (\#) & Triples (\#)\\
 \hline\hline
 Allie & 19,227,252 & 26 & 20,280,252 & 94,404,806 \\ 
 Cellcycle & 21,745 & 18 & 142,812 & 322,751\\
 DrugBank & 19,693 & 119 & 276,142 & 517,023 \\
LinkedSPL & 59,776 & 104 & 719,446 & 2,174,579 \\
 \hline
\end{tabular}
\caption{Characteristics of the KGs that were used to run the experiments along with detailed statistics depicted from columns 2-5. The first four columns show the KG name and its number of unique subjects, predicates, and objects, respectively. The last column depicts the total number of triples of each KG.}
\label{table::kgs}
\end{table}

\subsection{Knowledge Graph Benchmarks}
We used four well-known scientific KGs in this research. These are publicly available with a collection of relevant queries for each of the KGs. These KGs are also recognized as major KGs by previous studies such as~\cite{saleem,cellcycle,biobench}.

\textbf{Allie\footnote{\url{http://allie.dbcls.jp/}}} is a KG containing abbreviations and long forms utilized in life sciences. Allie contains all abbreviations and their corresponding long forms from titles and abstracts in the entire PubMed. \textbf{Cellcycle}\footnote{\url{ftp://ftp.dbcls.jp/togordf/bmtoyama/cellcycle/}} is a KG containing orthology relations for proteins. It consists of ten sub-graphs constituting the Cellcycle. In our experiments, we integrated them into a single KG dataset without modifying the content. \textbf{DrugBank\footnote{\url{https://download.bio2rdf.org/files/current/drugbank/drugbank.html}}} contains bioinformatics and chemoinformatics resource including detailed drug (chemical, pharmacological, pharmaceutical, etc.) and comprehensive drug targets such as sequence, structure, and pathway information.
\textbf{LinkedSPL\footnote{\url{https://download.bio2rdf.org/files/current/linkedspl/linkedspl.html}}} is already explained in the previous section. \reftab{table::kgs} shows the statistical information related to the above KGs.

\begin{table*}[h]
\centering
\begin{tabular}{cccccccccccc}
\hline
 \backslashbox{Benchmark}{Types}& Query & $SS^{a*}$ & $SO^{b*}$ & $Co^{c*}$ & $OPT^{d*}$ & Selective & $Fil^{e*}$ & $ORD^{f*}$ & $Lim^{g*}$ & $OFF^{h*}$ & $STP^{i*}$\\
 \hline\hline
\multicolumn{1}{ c  }{\multirow{5}{*}{Allie} } &
\multicolumn{1}{ c }{Q1} &  &  &  & & & & & & &  \cmark\\ 
\multicolumn{1}{ c  }{}                        &
\multicolumn{1}{ c }{Q2} &  &  &  & & & \cmark & & & &  \cmark\\ 
\multicolumn{1}{ c  }{}                        &
\multicolumn{1}{ c }{Q3} & \cmark &  &  & & &  & & & & \\ 
\multicolumn{1}{ c  }{}                        &
\multicolumn{1}{ c }{Q4} &  & \cmark  &  & & &  & & \cmark & & \\ 
\multicolumn{1}{ c  }{}                        &
\multicolumn{1}{ c }{Q5} & \cmark  & &  & & &  & \cmark & \cmark & & \\ \cline{1-12}
\hline
\hline
\multicolumn{1}{ c }{\multirow{6}{*}{Cellcycle} } &
\multicolumn{1}{ c }{Q1} &  &  & \cmark & & & & & & &  \\ 
\multicolumn{1}{ c }{}                        &
\multicolumn{1}{ c }{Q2} &  &  & \cmark & \cmark & & & & & &  \\ 
\multicolumn{1}{ c }{}                        &
\multicolumn{1}{ c }{Q3} &  &  & \cmark &  & \cmark & & & & &  \\ 
\multicolumn{1}{ c }{}                        &
\multicolumn{1}{ c }{Q4} &  &  & \cmark &  & & & & & &  \\ 
\multicolumn{1}{ c }{}                        &
\multicolumn{1}{ c }{Q5} & \cmark  &  &  & \cmark & & & & & &  \\ 
\multicolumn{1}{ c }{}                        &
\multicolumn{1}{ c }{Q6} & \cmark  &  &  &  & & & \cmark & & &  \\ \cline{1-12}
\hline
\hline
\multicolumn{1}{ c  }{\multirow{5}{*}{DrugBank} } &
\multicolumn{1}{ c }{Q1} & \cmark &  &  &  \cmark & & & & \cmark & &  \\ 
\multicolumn{1}{ c  }{}                        &
\multicolumn{1}{ c }{Q2} & \cmark &  &  & \cmark & &  & \cmark &\cmark &\cmark & \\ 
\multicolumn{1}{ c  }{}                        &
\multicolumn{1}{ c }{Q3} &  &  & \cmark & & &  & & & & \\ 
\multicolumn{1}{ c  }{}                        &
\multicolumn{1}{ c }{Q4} &  & \cmark  &  & & &  & & & & \\ 
\multicolumn{1}{ c  }{}                        &
\multicolumn{1}{ c }{Q5} &   & \cmark &  & & &  &  & \cmark & & \\ \cline{1-12}
\hline
\hline
\multicolumn{1}{ c  }{\multirow{2}{*}{LinkedSPL} } &
\multicolumn{1}{ c }{Q1} &  \cmark &   &  & & &  & &\cmark & & \\ 
\multicolumn{1}{ c  }{}                        &
\multicolumn{1}{ c }{Q2} &   & \cmark &  & & &  &\cmark  & \cmark &\cmark & \\ \cline{1-12}

\end{tabular}
\caption{Types of the queries. $SS^{a*}$: Subject-subject join, $SO^{b*}$: Subject-object join, $Co^{c*}$: combination of $SS$ and $SO$, $OPT^{d*}$: Optional pattern, $Fil^{e*}$: Filter, $ORD^{f*}$: Order by, $Lim^{g*}$: Limit, $OFF^{h*}$: Offset, $STP^{i*}$: Single triple pattern (no join)}

\label{table::queries}
\end{table*}

\textbf{Benchmark Queries.} KGs may contain four query forms, namely, SELECT, ASK, DESCRIBE, and CONSTRUCT. These forms are explained in the W3C portal in detail.\footnote{\url{https://www.w3.org/TR/sparql11-query/}} Similar to previous research such as~\cite{biobench,cellcycle,saleem}, our specific focus is on the SELECT queries in this paper. We selected 18 representative queries.\footnote{All queries are shown in Appendix A. All of them are also available through \url{https://github.com/oursubmission/SKG}} All or some of these queries have also been used in previous studies such as~\cite{biobench,cellcycle,saleem}. We ran these queries against the corresponding datasets using the DMSs. \reftab{table::queries} shows the classification of the 18 queries (see details of query types in the previous section).

\subsection{System Settings}
\label{sec:db}
\textbf{Computational Environment.} Our benchmark system was a physical machine with a 3.4GHz Corei7-3770 Intel processor, running Ubuntu Linux (kernel version: 4.15.0-88-generic), with 16GB of main memory, 8 cores, 256K L2 cache, 1TB instance storage capacity. The cache read is roughly 12865MB/sec and the buffer read is roughly 178.43MB/sec (the output of the ``hdparm -Tt'' Linux command). The operating system is set with almost no ``soft/hard'' limit on the file size, CPU time, virtual memory, locked-in-memory size, open files, processes/threads, and memory size.

\noindent \textbf{Data Management Systems (DMSs).} We chose four different DMSs as follows: (1) Virtuoso (Open Source Edition, Version 07.20.3230--commit a11a8e3), (2) Blazegraph\footnote{Previously known as Bigdata DB.} (Open Source Edition, version 2.1.6--commit 6b0c935), and MongoDB (Open Source Edition, version: 4.2.3). All or some of these DMSs have also been used in previous studies such as~\cite{watdiv,ISWC2013,Medha2,bsbm,biobench,saleem}.

\noindent \textbf{Configuration of Virtuoso.} We configured it based on the vendor's official recommendations.\footnote{\url{http://vos.openlinksw.com/owiki/wiki/VOS/VirtRDFPerformanceTuning}} We also used the latest version of GNU packages that are necessary to build Virtuoso (e.g. GNU gpref 3.0.4, libtool 2.4.6, flex 2.6.0, Bison 3.0.4, and Awk 4.1.3).

\noindent \textbf{Configuration of Blazegraph.} We configured Blazegraph based on the vendor's official performance tuning recommendations.\footnote{\url{https://wiki.blazegraph.com/wiki/index.php/PerformanceOptimization}} as well For example, we ran our experiments in the ``Worm'' standalone persistence store mode. We turned off all inference, truth maintenance, statement identifiers, and the free text index in our experiment since reasoning efficiency was not part of our research focus in this paper.

\noindent \textbf{Configuration of RDF-3X and MongoDB.} We used the default settings for both RDF-3X and MongoDB.

\noindent \textbf{Indexing of Virtuoso.} We did not change the default indexing scheme of Virtuoso. As highlighted in the official website, ``alternate indexing schemes are possible but will not be generally needed\footnote{\url{http://docs.openlinksw.com/virtuoso/rdfperfrdfscheme}}''. Virtuoso creates the following compound indexes by default for the loaded KGs: PSOG, POGS, SP, OP, and GS.

\noindent \textbf{Indexing of Blazegraph.} As recommended in the Blazegraph's official website,\footnote{\url{https://wiki.blazegraph.com/wiki/index.php/PerformanceOptimization}} we did not change its default data modeling or the indexing schema. 

\noindent \textbf{Indexing of RDF-3X.} It creates exhaustive indexes over a single ``giant triples table'' by building indexes over all six permutations of the three dimensions that constitute an RDF triple, and additionally, indexes over count-aggregated variants for all three two-dimensional and all three one-dimensional projections.

\noindent \textbf{MongoDB Indexing.} We created indexes on those name/value pairs of the JSON representations that were representatives of subjects and predicates.

\noindent \textbf{Loading the scientific KGs.} We loaded the RDF/N-Triples format of KGs into \textit{Virtuoso} by using its native bulk loader function (``ld\_dir''). To load the KGs into \textit{Blazegraph}, we used Blazegraph's native ``DataLoader'' utility.\footnote{\url{https://wiki.blazegraph.com/wiki/index.php/Bulk_Data_Load}}Since JSON-LD has gradually become the de-facto standard to represent RDF datasets in JSON formats,\footnote{\url{https://www.w3.org/2018/jsonld-cg-reports/json-ld/}} we converted the KG datasets from RDF/N-Triples syntax to JSON-LD using a parser designed and developed as part of this project\footnote{The source code is available through \url{https://github.com/oursubmission/SKG}} to load them into MongoDB using its native tool called ``mongoimport''. In a similar way, we used RDF-3X's native tool called ``rdf3xload'' to load KGs.

\noindent \textbf{Shutdown store, clear caches, restart store.} We measured the query execution times in our evaluation. This is an end-to-end time computed from the time of query submission to the time when the result is outputted. After the execution of each query, we carefully checked to ensure that the output results are correct and exactly the same across different DMSs. The query times for both cold- and warm-run (aka, cold and warm cache) are reported. For cold-run we dropped the file systems caches using \mbox{\texttt{echo 3 > /proc/sys/vm/drop\_caches}} and \mbox{\texttt{swapoff -a}} commands. For fairness, the warm-run query times reported for each DMS are averaged (\textit{geometric mean}) over 5 successive runs (with almost no delay in between) to account for any randomness and noise.

\begin{figure*}
        \centering
        \begin{subfigure}[b]{0.375\textwidth}
            \centering
            \includegraphics[width=\textwidth]{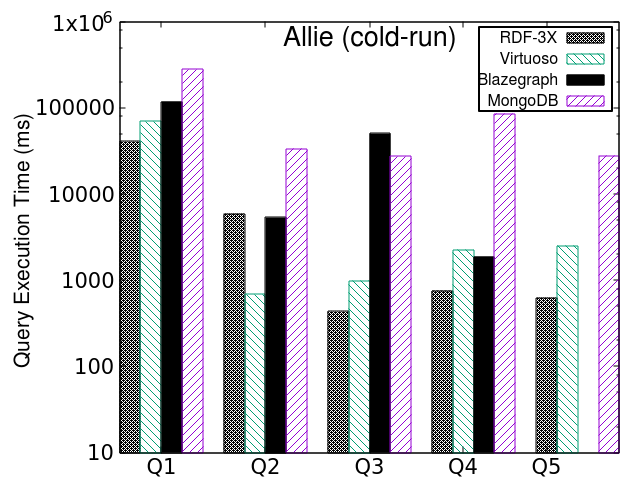}
            \caption[]%
            {{}}    
            \label{cold-allie}
        \end{subfigure}
        \quad
        \begin{subfigure}[b]{0.375\textwidth}  
            \centering 
            \includegraphics[width=\textwidth]{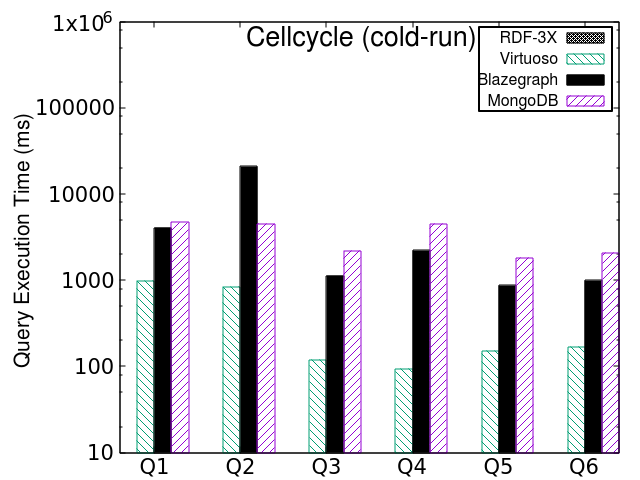}
            \caption[]%
            {{}}    
            \label{cold-cell}
        \end{subfigure}
        \vskip\baselineskip
        \begin{subfigure}[b]{0.375\textwidth}   
            \centering 
            \includegraphics[width=\textwidth]{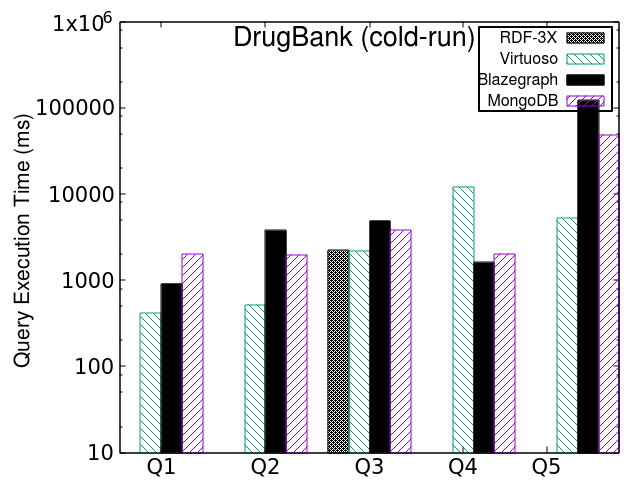}
            \caption[]%
            {{}}    
            \label{cold-drug}
        \end{subfigure}
        \quad
        \begin{subfigure}[b]{0.375\textwidth}   
            \centering 
            \includegraphics[width=\textwidth]{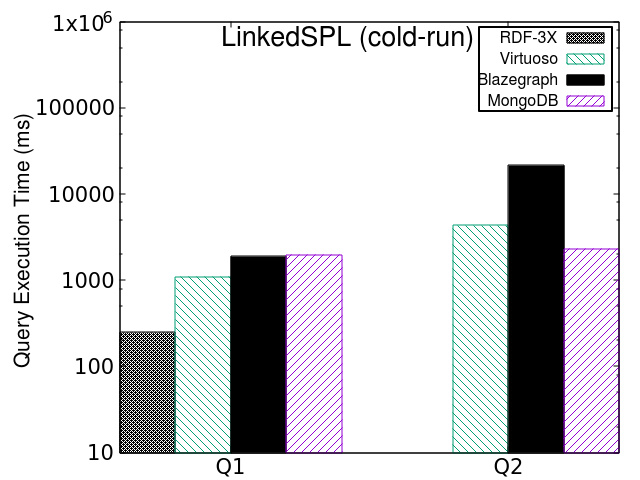}
            \caption[]%
            {{}}    
            \label{cold-link}
        \end{subfigure}
        \caption[ The average and standard deviation of critical parameters ]
        {\small The cold-run execution times of different queries against each KG. The $X$ axis shows different queries. The $Y$ axis shows the execution time of each query in milliseconds (\textbf{log scale}). No value is shown when a query is not supported by a DMS (e.g., RDF-3X does not support queries of Cellcycle KG) or the returned result is different with others (e.g., Blazegraph's result for Allie-Q5).} 
        \label{cold-result}
    \end{figure*}

\section{Evaluation}
\label{sec::Evaluation}
\noindent We evaluated the query performance of the DMSs for scientific KG. Our goal is to discover and explain systematic performance differences, if any.

\subsection{Results}
\label{sec::results}
\begin{figure*}
        \centering
        \begin{subfigure}[b]{0.375\textwidth}
            \centering
            \includegraphics[width=\textwidth]{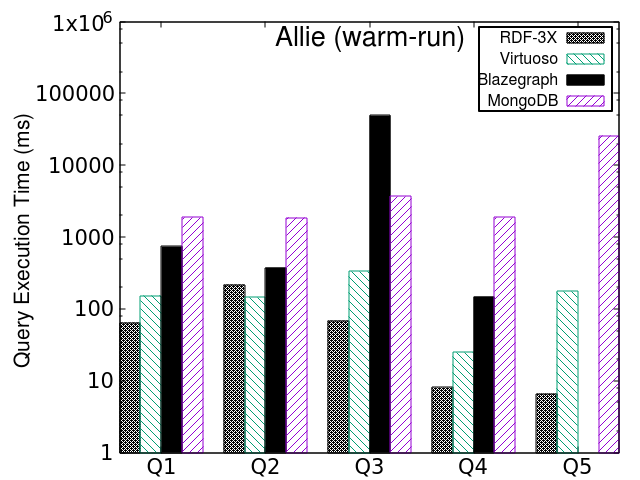}
            \caption[]%
            {{}}    
            \label{warm-allie}
        \end{subfigure}
        \quad
        \begin{subfigure}[b]{0.375\textwidth}  
            \centering 
            \includegraphics[width=\textwidth]{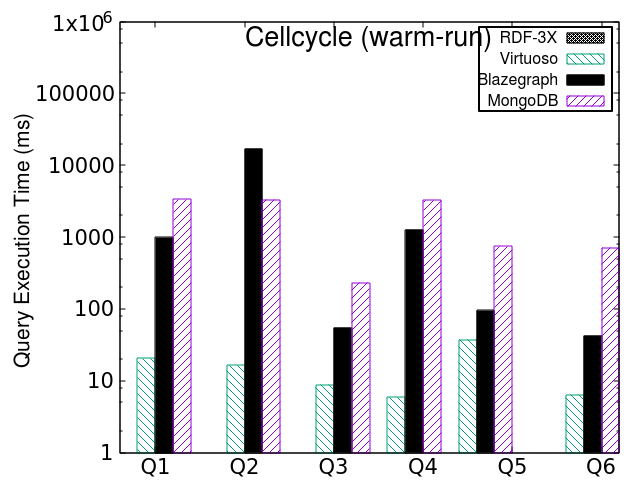}
            \caption[]%
            {{}}    
            \label{warm-cell}
        \end{subfigure}
        \vskip\baselineskip
        \begin{subfigure}[b]{0.375\textwidth}   
            \centering 
            \includegraphics[width=\textwidth]{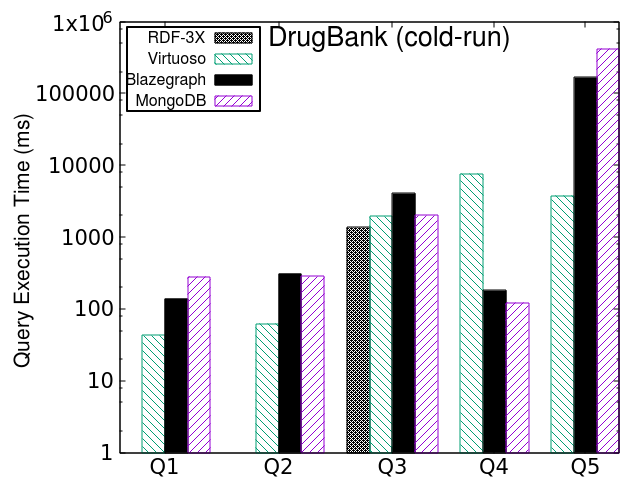}
            \caption[]%
            {{}}    
            \label{warm-drug}
        \end{subfigure}
        \quad
        \begin{subfigure}[b]{0.375\textwidth}   
            \centering 
            \includegraphics[width=\textwidth]{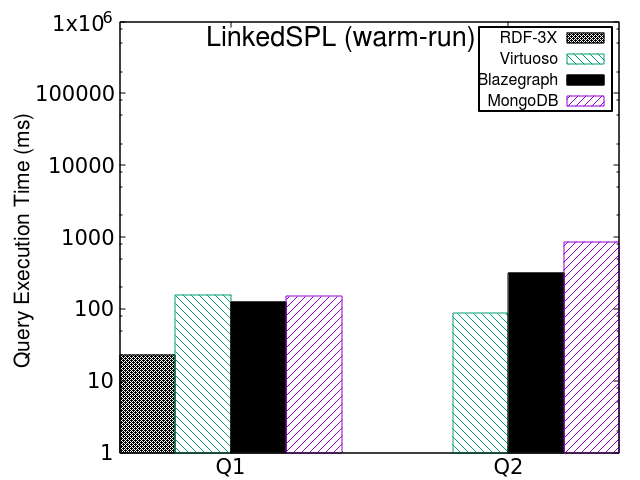}
            \caption[]%
            {{}}    
            \label{warm-link}
        \end{subfigure}
        \caption[ The average and standard deviation of critical parameters ]
        {\small The warm-run execution times of different queries against each KG. The $X$ axis shows different queries. The $Y$ axis shows the execution time of each query in milliseconds (\textbf{log scale}). No value is shown when a query is not supported by a DMS (e.g., RDF-3X does not support queries of Cellcycle KG) or the returned result is different with others (e.g., Blazegraph's result for Allie-Q5).} 
        \label{warm-result}
    \end{figure*}

The query execution times over the KGs are presented in \reffig{cold-result} (cold-run) and \reffig{warm-result} (warm-run). In these figures, the $X$ axis shows the different queries and the $Y$ axis shows the execution times of queries in milliseconds (using \textbf{log scale}). Note that RDF-3X cannot support queries with complex triple patterns, filtering, offset modifiers, and optional patterns like queries of the Cellcycle KG (e.g., \reffig{cold-cell}). In these cases, no value is shown for RDF-3X.

The cold-run results suggest that RDF-3X offers several orders of magnitude performance advantages over others for queries with a single triple pattern (i.e., no join) and less complex triple patterns (e.g., no optional or complex filtering patterns) such as Allie-Q1 and Allie-Q3-5 (\reffig{cold-allie}). However, this DMS could not execute Allie-Q2 as fast as others since this query contains a filtering pattern. Virtuoso exhibits around one order of magnitude better performance to run complex queries which are queries with a combination of subject-to-subject and subject-to-object joins (see \reffig{cold-cell}). Blazegraph showed relatively better performance to execute subject-to-object join queries like DrugBank-Q4 or Allie-Q4 (as compared to Virtuoso). MongoDB as a document-store could execute all the queries. For subject-to-subject join queries like DrugBank-Q1-2 or LinkedSPL-Q1, its performance is comparable with others. However, MongoDB did not display consistently good performance in our experiment. In the warm-run results (\reffig{warm-result}), the trends related to the performance of different DMSs are almost remained unchanged. These results show that there are interactions between different query types and DMSs. In an attempt to explain the factors contributing to the performance differences, we present our detailed analyses for each DMS with regard to different query types below.

\subsection{Analysis}
\label{sec::results}

\textbf{RDF-3X.} This DMS creates exhaustive indexes on all permutations of triples, their binary, and unary projections. These indexes are compressed to reduce storage space and fit better in the main memory. Its query processor is designed to aggressively leverage cache-aware hash and merge joins. In addition, query optimization typically requires selectivity estimation, RDF-3X uses a single-dimensional histogram to estimate data statistics. These design principles enable RDF-3X to scan relevant indexes for each triple pattern of a query separately to retrieve the result of each triple pattern efficiently. RDF-3X keeps the retrieved result in the main memory as an intermediary result and then aggressively uses a merge join algorithm to join the intermediary results and return the final result. RDF-3X's workload-independent, exhaustive indexing alongside the implementation of a merge join algorithm is the most probable reason behind its performance advantage for executing queries on Allie KG. However, RDF-3X's query optimization relies on join ordering using dynamic programming for plan enumeration with a statistics-based cost model. It means that the existence of modifiers like optional patterns or filtering expressions in a query is an anti-pattern for the RDF-3X query processor and affects its performance negatively. For instance, Allie-Q2 has just a simple filtering expression more than Allie-Q1, which is enough to affect the performance superiority of RDF-3X negatively.

\noindent \textbf{Virtuoso.} Similar to RDF-3X, Virtuoso's physical design is based on a relational table with three columns\footnote{In the case of loading named graphs, it adds another column for the context, called C.} for S, P, and O (S: Subject, P: Predicate, and O: Object) and carrying multiple bitmap indexes over that table to provide a number of different access paths. Most recently, Virtuoso added columnar projections to minimize the on-disk footprint associated with RDF data storage. After constructing the underlying table, Virtuoso translates any input SPARQL to the equivalent SQL and then executes the SQL over the table. To run the SQL queries faster, Virtuoso uses indexed-loop joins with optional inline checks for filter expressions, hash joins, and bloom filters (when applicable). These techniques enable Virtuoso to execute combined queries (a combination of subject-to-subject and subject-to-object join queries) faster than other DMSs. Virtuoso also implemented some techniques for joins on matching beginnings of keys in the intermediary results to implemented a merge-join-like operation, but it is technically an indexed-loop join with frequent checking alongside with intermediary results pruning. Based on that, the most probable reason behind, Virtuoso's slower execution of subject-to-object join queries is the lack of merge join implementation. Virtuoso parses SPARQL queries efficiently into internal tree-like representation and other compiler's internal data at the same time, however, this translation process imposes some run-time overhead to the query processing which is probably the most important reason behind slower execution of Allie-Q4 or DrugBank-Q4. Virtuoso also caches recently accessed parts of source data in tables, as well as, the compiled queries (results will not be cached) which may contribute to the performance improvement of Allie-Q4 in the warm-run.

\noindent \textbf{Blazegraph.} Blazegraph's physical design is based on \mbox{B+Trees} to store KGs in the form of ordered data. Blazegraph typically uses the following three indexes based on the stored \mbox{B+Trees} for triples modes: SPO, POS, and OSP. For normal use cases, these indexes are laid out on variable-sized pages. These index pages are read from the backing store and load in the main memory on demand into the Java heap. However, Blazegraph takes advantage of a variety of data structures to execute queries when stored the content is loaded in the main memory. For example, the underlying data model (i.e., \mbox{B+Trees}) is retained by a mixture of a ring buffer (hard reference queue), weak references, and hard references on the stack during the use alongside a native memory cache for buffering writes to reduce write application effects. Blazegraph typically uses either a hash join and index-nested-loop join to support joins in the run-time over the intermediary results. Although it does not use merge join for subject-to-object join queries, the effectiveness of Blazegraph \mbox{B+Trees} implementation shows itself by a faster execution of Allie-Q4 as compared to the bitmap indexing of Virtuoso. In addition to the effective implementation of the \mbox{B+Trees}, a better cardinality estimation may contribute to the performance advantages of Blazegraph for executing subject-to-object join as compared to Virtuoso.

\noindent \textbf{MongoDB.} MongoDB uses a key/value store as an internal storage engine (i.e., by default, WiredTiger) to store JSON documents\footnote{MongoDB uses the binary equivalent of each JSON document (i.e., BSON) for storage, in which the structure of each document remained unchanged}. It usually assigns an arbitrary (and unique) identifier to each JSON document as a key and considers the document as a value to store them. MongoDB uses \mbox{B-Trees} to create indexes on the contents of each JSON document. By using JSON-LD representation, all triples with the same subject have appeared in a single JSON document and with a constructed index on subjects, the joining of triples with the same subject is equivalent to an index-based look-up querying of a given subject. Therefore, we typically expect to observe better performance from MongoDB for subject-subject join queries. However, our experiments showed that MongoDB may not outperform other DMSs even for subject-to-subject joins. It appears to be because of low-selectivity and analytical nature of scientific queries where even for subject-to-subject join queries MongoDB cannot take advantage of its indexes. In other words, in our experiments, execution of each scientific query against MongoDB was tended to be almost a full-scan of the entire corresponding KG which contributed to the slower execution times for MongoDB as compared to other DMSs. In addition to the low-selectivity of scientific queries, MongoDB just uses index-nested-loop join for query processing which usually contributes to slower execution time for scientific queries. Allie-Q4 (i.e., subject-to-object join query) is a good example to display the efficiency of different join algorithms where RDF-3X with a merge join implementation is faster than Blazegraph with a hash join implementation based on \mbox{B+Trees} indexes and Virtuoso is slower than these two DMSs with its bitmap-index-based hash join and MongoDB with an index-nested-loop join is the slowest to execute Allie-Q4.

\section{Discussion}
\label{sec::discussion}
Our experimental evidence showed that the variability in the data and query requirements of the scientific domain are less likely to be matched by even the most sophisticated (single) state-of-the-art DMS. A closer look at the state-of-the-art reveals that they trace their roots to techniques and architectures from the 1970s~\cite{rajbook,theEnd,rethink,rethinkrevisit,Dan2016short}. For example, Virtuoso borrowed heavily from object-relational systems, RDF-3X and Blazegraph implemented different variations of the legacy B-tree data structure to create exhaustive indexes for KGs. In general, the first releases of almost all current tools were architected around 20 years ago typically based on the following design choices (i) disk-oriented persistent model, (ii) disk-resident indexes, (iii) write-ahead logging for recovery, (iv) multi-threading and buffer-pooling to reduce latency, etc~\cite{rajbook,theEnd}. There have been a number of extensions over the past years, ranging from supporting compression and columnar storage to bitmap indexes. Similarly, vectored execution, compiler-based early evaluation of query expressions (i.e., data-independent sub-queries), clustering of the underlying data based on workload patterns, and run-time join type selection (e.g., merge or hash joins), to name a few, are among other major extensions which have successfully been implemented. However, due to the unprecedented challenges in terms of heterogeneity (variety) and velocity of KG data, it appears to be that a single state-art-of-the DMS is unlikely to be able to manage the \textit{heterogeneity} of data formats and to optimize the \textit{performance} of data accesses. As well, the individual \textit{queries} executed over KGs have become \textit{highly diverse}. The upshot has been that the behavior and \textit{performance} of data analysis have become \textit{unpredictable}.

To address the unprecedented performance challenges of scientific KG query processing, we briefly outline two key approaches: (i) a complete redesign of RDF-stores adopting NVM-oriented architectures (Non-Volatile Memory) with query compilation and (ii) architecturing a multi-database system that can offer a genuine polygloty at the level of query and access languages and data persistence. This approach is inspired by Ashby's First Law of Cybernetics~\cite{ashby} which can be paraphrased in this context to state that the variety in the solution architecture should be greater than or at least equal to that of the variety displayed by the data and the queries. We propose that the requisite variety can be achieved through an architecture based on the emerging hardware devices like NVMs or/and by providing a \textit{polyglot} model of data persistence supported by an \textit{intelligent workload management} design that can analyze individual queries and match each to the combination of likely best-performing persistent store and database engine. These approaches are briefly explained below.

\subsection{NVM-oriented RDF-stores}
The emergence of NVM has fundamentally changed the dichotomy between dynamic RAM and hard drive storage~\cite{rajbook}. NVM devices are almost as fast as dynamic memories, but their data remained persistent even after power loss. One of the lessons we have learned from our study is that current RDF-stores are built under the assumption that memory is volatile and the data needs to be loaded into memory from the disk whenever requested by queries. However, an NVM-based RDF-store can avoid this indirection and store the data and indexes (i.e., usually direct pointers to records) only in NVM for performance gain. In addition, we note that current RDF-stores typically use different disk-oriented iterator-based query processing models (inspired by Volcano-style processing~\cite{volcano}) which impose some overhead and increase query execution times~\cite{DBS-mainmemory}, but this overhead will be eliminated by direct compilation of queries into low-level machine codes and run them directly over the records. In short, SPARQL query processing has two major stages: (1) parsing query language grammar, generating the corresponding syntax tree, and transforming the syntax tree into an optimized logical operator graph and (2) choosing the best implementations for each logical operator (also referred to as physical optimization). Our argument about the direct compilation of queries targeted the second stage of SPARQL query processing since the first stage is almost the same in disk-based and NVM-oriented RDF-stores. An efficient DMS needs to use the minimum number of instructions to implement physical plans of queries by either writing code that converts a SPARQL query plan to C/C++ and then compile and run it to generate native code or compiling a SPARQL query directly to a corresponding native code using LLVM\footnote{\url{https://llvm.org/}} toolkit~\cite{llvm,compindiana,compilebook}. Adopting each of these two compilation techniques can affect the efficiency of SPARQL query processing significantly.


\subsection{Genuine Polygloty}
Our approach is guided by the conclusion that a single, one-size-fits-all DMS~\cite{oneSize,onesize2} is unlikely to emerge and that the critical research task is to achieve cross-platform integration whereby ``$\dots$platforms will need to be \textit{integrated} or \textit{federated} to enable data analysts and analyze data \textit{across systems}''~\cite{Dan2016short}. Over the past few years, there has been growing interest in employing multiple DMSs (as opposed to ``one size fits all'' strategy) for processing data-intensive applications with diverse requirements~\cite{lim2013}. Such interest led to the development of some open-source platforms such as Apache Beam\footnote{\url{https://beam.apache.org}} and Drill\footnote{\url{https://drill.apache.org}} as well as some academic prototypes such as~\cite{bigdawg}. In general, these projects proposed multi-database systems to support multiple data models against a single, integrated backend that can potentially address the growing requirements for scalability and performance~\cite{lookliu}. However, the focus of current polygloty solutions is on applications such as OnLine Analytical Processing (OLAP) and rather less attention has been paid to scientific query processing. As well, it is not difficult to see that the lack of integration across the entire data will lead to balkanized data islands that cannot support applications that cut across the separate data stores.

We seek to achieve polygloty at both the access and persistence layers with the ability of matching the query requirements with the best combination of DMS and storage representation to achieve improved query execution performance is in the true spirit of polygloty. This approach has the potential to achieve the requisite variety that is needed to query scientific data efficiently. The proposed approach includes three layers: scientific applications, intelligent workload management, and polyglot persistence. A scientific application interacts with the approach like it interacts with any conventional single DMS. For example, the application may send their workload issuing a declarative query language like SPARQL. Traditional DMSs usually consist of one execution engine and one storage engine where these two engines are tightly-coupled and cannot perform individually. On the contrary, the proposed architecture must contain multiple DMSs internally where the intelligent workload management layer has the responsibility of selecting one or more of the employed DMSs that can best serve requests made by each application. The workload management layer needs to directly use the execution and/or storage engines of the underlying DMSs in the polyglot persistence layer. This enables it to have full control of what gets executed and how. For instance, the polyglot persistence layer could consist of four DMSs that we employed in this study, namely, Virtuoso, Blazegraph, RDF-3X, and MongoDB.

\section{Related Work}
\label{sec::related_work}
There is growing interest in the use of KGs available in the form of RDF datasets for the representation, exchange, and reuse of scientific data processing. Data management of RDF datasets has been the research focus of several studies so far. Early approaches such as ~\cite{Broek3,Jena17,AntiDan} employed relational database systems to store them. These systems typically store a set of triples by using a relational table with three columns resulting in low implementation overhead. Virtuoso~\cite{Virtuso} and RDF3X~\cite{RDF3x} are well-known systems from this category. Abadi et al.~\cite{Dan2007,DanSW} represented some of the first studies in which the importance of data representation using \textit{SQL-based} systems was highlighted and the use of column-oriented DMSs (e.g.,~\cite{CStore}) was proposed. Over time, the emergence (and the growing use) of KGs called for systems that can store and evaluate \textit{queries} over them efficiently~\cite{Growing3,Growing2,Growing1,Survey2018}. In response, a variety of DMSs were proposed such as Blazegraph. As discussed in comprehensive surveys such as~\cite{Kaoudi,Ozsu,Survey2018}, we can classify the previous studies into several categories. We briefly review three major categories, namely, triple-based indexing, infrastructure configuring, and graph processing in the following.

\textbf{Triple-based Indexing.} Virtuoso, HexaStor~\cite{Hexastore}, and Rya system~\cite{Rya}, to name a few, are three DMSs that are performing mainly based on indexing. For instance, the Rya~\cite{Rya} which is designed on the top of Accumulo~\footnote{https://accumulo.apache.org/} (i.e., a distributed key-value and column-oriented NoSQL store) created indexes on the all permutations of the triple pattern across three separated tables. The permutations include SPO (S stands for Subjects, P stands for Predicates, and O stands for Objects), POS, and OSP. The effectiveness of triple-based indexing solutions can be limited since querying KGs typically requires touching a large amount of data and complex filtering.

\textbf{Infrastructure Configuring.} JenaHBase~\cite{JenaHbase}, H2RDF~\cite{H2rdf}, and AMADA~\cite{AMADA} are three well-known DMSs that focused mainly on the importance of configurations of underlying infrastructure such as cluster segmentation, communication overhead, and distributed storage layouts. For instance, JenaHBase~\cite{JenaHbase} proposed a custom-built data storage layout for query processing and physical storage.

H2RDF~\cite{H2rdf} combines the HBase\footnote{https://hbase.apache.org/} and the Hadoop\footnote{https://hadoop.apache.org/} framework. H2RDF employed the Hadoop platform to provide a distributed query processing module by launching MapReduce jobs for queries that require touching a large amount of data. H2RDF+~\cite{H2rdfp} extended the H2RDF~\cite{H2rdf} by creating indexes on all permutations of triple patterns in distributed indexing tables. In other words, H2RDF+~\cite{H2rdfp} merged triple-based indexing and infrastructure configuration techniques.

AMADA~\cite{AMADA} also exploited infrastructure configuration techniques by employing cloud computing to store and query data. In particular, AMADA stores the data in the Amazon Simple Storage Service (S3). The S3 interface attaches a URL to each dataset to be used later for the query processing. AMADA used Amazon Simple Queue Service (SQS) and virtual machines within the Amazon Elastic Compute Cloud (EC2) for the query execution.

\textbf{Graph Processing.} Some approaches have applied ideas from the graph processing world to handle KG querying such as Blazegraph, gStore~\cite{gStore}, and \cite{Turbo}. For instance, gStore~\cite{gStore} as a graph-based storage system models KGs as a labeled and directed multi-edge graph. gStore stores the graph by using a disk-based adjacency list table and executes queries by mapping them to a subgraph matching task over the graph. Kim et. al.~\cite{Turbo} considers RDF graphs as labeled graphs and applies subgraph homomorphism methods for query processing. To improve its query performance, it exploits optimization techniques and a Non-Uniform Memory Access (NUMA)-aware parallelism for query processing.

In addition to the design of the DMSs, analysis of available DMSs using benchmark datasets has been a core topic of data management research. For example, some studies such as~\cite{bsbm,watdiv}, to name a few, presented new benchmark datasets. Some other studies such as~\cite{ISWC2013} did not propose any new dataset but tried to use available benchmarks and DMSs for reporting key advantages and drawbacks of each DMS. There are also studies such as~\cite{saleem} which surveyed and analyzed available datasets in terms of different metrics such as the number of projection variables, the number of BGPs, etc. In contrast to these studies, our particular focus is to provide a fine-grained, comparative performance analysis of the major DMS types against scientific KGs.

\section{Conclusion}
\label{sec::conclusion}
The increase in the heterogeneity of scientific datasets and the growing interest in the use of KGs for the representation, exchange, and reuse of these datasets have triggered the development of a range of DMSs broadly classified as document, columnar, and graph stores in addition to the relational. In this paper, we have provided experimental evidence to show that the variability in the scientific data and query requirements cannot be matched by even the most sophisticated state-of-the-art (single) DMS. We have addressed some of the critical performance challenges associated with these platforms in the context of KGs by briefly outlining two key approaches: (1) a complete redesign of RDF-stores adopting NVM-oriented architectures and direct query compilation and (2) architecturing a genuine polygloty at the level of query and access languages and data persistence. We have argued that an NVM-oriented RDF-store can avoid the disk-based data and indexes retrieval and execute SPARQL queries against KGs by converting them into optimized low-level executable machine codes for significant performance gain. We have also discussed an architecture that can achieve genuine polygloty at the level of access languages and data persistence to classify queries, analyze individual query types and match each to the best performing platform. Further steps also include efforts to minimize the amount of data replications without negatively affecting the robustness and performance. We are in the process of implementing and experimenting with prototype systems based on the approaches outlined in this paper.

\bibliographystyle{ACM-Reference-Format}
\bibliography{main}


\vfill\null

\newpage

\appendix

\section*{Appendix}
\section{SPARQL Queries}
\label{appQ}
For completeness we include the SPARQL queries used in our evaluation.

\noindent \textbf{Allie.} This KG came with 5 SPARQL queries.\footnote{https://hobbitdata.informatik.uni-leipzig.de/benchmarks-data/queries/biobench-allie-queries.txt} We simplified the queries to quantify how fast the DMSs can run queries with the minimum number of triple patterns. For instance, no join was required to execute Allie-Q1 or Allie-Q2 and the rest of them mainly needed to perform subject-subject joins to return the results.

\noindent \textbf{Allie-Q1:} 
\begin{verbatim}
select * 
where {
?s allie:inResearchAreaOf ?X .
}
\end{verbatim}

\noindent \textbf{Allie-Q2:} 
\begin{verbatim}
select *
where  { 
?s allie:inResearchAreaOf ?X .
filter ( ?X = <%p%> )
}
\end{verbatim}

\noindent\textbf{Allie-Q3:} 
\begin{verbatim}
select *
where {
?s allie:inResearchAreaOf ?X .
?s rdfs:label ?y . }
\end{verbatim}

\noindent\textbf{Allie-Q4:} 
\begin{verbatim}
select *
where  {
?s allie:hasMemberOf ?x .
?x <http://purl.org/allie/ontology/201108#frequency> ?o.
} limit 10
\end{verbatim}

\noindent\textbf{Allie-Q5:} 
\begin{verbatim}
select *
where  {
?s allie:appearsIn ?x10 .
?s allie:cooccursWith ?x2.
?s allie:frequency ?x3.
?s allie:inResearchAreaOf ?x4.
?s allie:hasLongFormOf ?x5.
?s allie:hasShortFormOf <%p%>.
?s rdfs:type ?x7 .
?s allie:appearsIn ?x8.
?s allie:cooccursWith ?x9.
} order by ?x4 limit 100
\end{verbatim}

\vfill\null

\newpage

\noindent \textbf{Cellcycle.} We chose 6 analytic complex queries\footnote{We selected these queries from the following source: https://hobbitdata.informatik.uni-leipzig.de/benchmarks-data/queries/cell.biobench.queries.txt} to run against Cellcycle KG. These queries are mainly low-selective with subject-object joins requirement.

\noindent \textbf{Cellcycle-Q1:} 
\begin{verbatim}
select *
where {
?protein_id   ssb:has_function    ?function_id.
?function_id  ssb:is_a     <%p1%>.
?protein_id   ssb:located_in      ?location_id.
?location_id  ssb:is_a     <%p2%>.
?protein_id   ssb:participates_in  ?process_id.
?protein_id   rdfs:label              ?protein. }
\end{verbatim}

\noindent \textbf{Cellcycle-Q2:} 
\begin{verbatim}
select *
where {
?protein_id rdf:type        ssb:protein.
?protein_id ssb:Definition  ?Def.
?Def        ssb:def         ?definition.
?protein_id rdfs:label      ?protein_name.
OPTIONAL {
?protein_id  ssb:participates_in   ?interaction.
?interaction rdf:type           ssb:interaction.
?interaction rdfs:label       ?interaction_name.
?interaction ssb:xref                     ?xref.
?xref        ssb:acc                 ?IntAct_id.
}} 
\end{verbatim}

\noindent \textbf{Cellcycle-Q3:} 
\begin{verbatim}
select ?description ?transformed_protein_name ?cco_id
where  {
<%p%>  ssb:transforms_into    ?cco_id.
?cco_id      ssb:Definition      ?Def.
?Def         ssb:def     ?description.
?cco_id      rdfs:label ?transformed_protein_name.
}
\end{verbatim}

\noindent \textbf{Cellcycle-Q4:} 
\begin{verbatim}
select *
where  {
?protein_id      rdf:type            ?o1.
?protein_id      ssb:is_a            ?o2.
?protein_id      ssb:has_function    ?subfunction_id.
?subfunction_id  ssb:is_a            ?function_id.
?protein_id      ssb:located_in      ?location_id.
?location_id     ssb:is_a            <%p%>.
?function_id     ssb:Definition      ?def.
?def             ssb:def        ?function.
?protein_id      rdfs:label  ?protein_name.}  
\end{verbatim}

\vfill\null

\newpage

\noindent \textbf{Cellcycle-Q5:} 
\begin{verbatim}
select *
where  {
?term_id      ssb:has_source        <%p%> .
?term_id      ssb:participates_in  ?interaction.
?term_id      rdf:type             ?type.
optional{
?term_id      rdfs:label           ?protein.
}} 
\end{verbatim}

\noindent \textbf{Cellcycle-Q6:} 
\begin{verbatim}
select *
where  {
?term_id      ssb:has_source        <%p%> .
?term_id      ssb:participates_in  ?interaction.
?term_id      rdf:type             ?type.
?term_id      rdfs:label           ?protein.
}  order by desc(?protein)
\end{verbatim}

\noindent \textbf{DrugBank.} We formulated 5 SPARQL queries to quantify the DMSs' efficiency to execute queries with a complex mixture of join and optional patterns, as well as, filtering expressions in disjunctive form and modifiers like ``order by'' and ``offset''.

\noindent \textbf{DrugBank-Q1:} 
\begin{verbatim}
select ?drug_uri ?label ?indication ?mechanismOfAction
?biotransformation ?halfLife
where {
?drug_uri a drugbank:drugs .
?drug_uri rdfs:label ?label .
OPTIONAL { ?drug_uri drugbank:brandName ?brandName . }
OPTIONAL { ?drug_uri drugbank:indication ?indication . }
OPTIONAL { 
?drug_uri drugbank:mechanismOfAction ?mechanismOfAction .
}
OPTIONAL { 
?drug_uri drugbank:biotransformation ?biotransformation .
}
OPTIONAL { ?drug_uri drugbank:halfLife ?halfLife . }
} limit 100
\end{verbatim}

\noindent \textbf{DrugBank-Q2:} 
\begin{verbatim}
select ?drug_uri ?label ?indication ?mechanismOfAction
?biotransformation ?halfLife
where {
?drug_uri a drugbank:drugs .
?drug_uri rdfs:label ?label .
OPTIONAL { ?drug_uri drugbank:brandName ?brandName . }
OPTIONAL { ?drug_uri drugbank:indication ?indication . }
OPTIONAL {
?drug_uri drugbank:mechanismOfAction ?mechanismOfAction .
}
OPTIONAL {
?drug_uri drugbank:biotransformation ?biotransformation .
}
OPTIONAL { ?drug_uri drugbank:halfLife ?halfLife . }
} order by ?label desc(?brandName) ?mechanismOfAction offset 100 limit 100
\end{verbatim}

\vfill\null

\newpage

\noindent \textbf{DrugBank-Q3:} 
\begin{verbatim}
select ?Drug ?IntDrug ?IntEffect
where {
?y <http://www.w3.org/2002/07/owl#sameAs> ?Drug .
?Int drugbank:interactionDrug1 ?y .
?Int drugbankinteractionDrug2 ?IntDrug .
?Int drugbank:text ?IntEffect .
} 
\end{verbatim}

\noindent \textbf{DrugBank-Q4:} 
\begin{verbatim}
select  *
where {
?s  drugbank:interactionDrug1 ?o.
?s  drugbank:interactionDrug2 ?o5.
?s  drugbank:text ?o6.
?o  drugbank:biotransformation ?o2 .
?o  drugbank:brandName ?o3.
?o  drugbank:ahfsCode ?o4 .
?o  drugbank:absorption ?o7 .
?o  drugbank:affectedOrganism ?o9 .
?o  drugbank:brandMixture ?o10 .
?o  drugbank:atcCode ?o11 .
?o  drugbank:casRegistryNumber ?o12.
?o  drugbank:chemicalFormula ?o13 .
?o  drugbank:meltingPoint ?o14 .
?o  owl:sameAs ?o15 .
filter ( ?o11 != <%p1%> || !(?o11 = <%p2%>) )
} limit 200
\end{verbatim}

\noindent \textbf{DrugBank-Q5:} 
\begin{verbatim}
select *
where {
?s  drugbank:interactionDrug1 ?o.
?s  drugbank:interactionDrug2 ?o5.
?s  drugbank:text ?o6.
?o  drugbank:biotransformation ?o2 .
?o  drugbank:brandName ?o3.
?o  drugbank:ahfsCode ?o4 .
?o  drugbank:absorption ?o7 .
?o  drugbank:affectedOrganism ?o9 .
?o  drugbank:brandMixture ?o10 .
?o  drugbank:atcCode ?o11 .
?o  drugbank:casRegistryNumber ?o12. 
?o  drugbank:chemicalFormula ?o13 .
?o  drugbank:eltingPoint ?o14 .
?o  owl:sameAs ?o15 .
?o15 ?p3 ?o8 .
} limit 10
\end{verbatim}

\vfill\null

\newpage

\noindent \textbf{LinkedSPL.} We formulated 2 SPARQL queries for this KG which are inspired by sample queries of Medical SPARQL Query Library.\footnote{https://www.w3.org/wiki/HCLSIG/Use\_case/Medical\_SPARQL/queries}

\noindent \textbf{LinkedSPL-Q1:}
\begin{verbatim}
select *
where {
?s linkedSPL:activeMoietyRxCUI ?x6 .
?s linkedSPL:pharmgxBiomarker ?x1 .
?s linkedSPL:pharmgxDrug ?x2 .
?s linkedSPL:pharmgxSPLSection ?x5 .
?s linkedSPL:pharmgxXref ?x4 .
?s linkedSPL:setId ?x3 .
?s linkedSPL:therapeuticApplication ?x7 .
?s rdf:type linkedSPL:pharmgxData .
?s owl:sameAs ?x8 .
} limit 100
\end{verbatim}

\noindent \textbf{LinkedSPL-Q2:} 
\begin{verbatim}
select  *
where {
?s   rdfs:label            ?x .
?s   linkedSPL:howSupplied  ?x5 .
?s   linkedSPL:adverseReactions  ?x2 .
?s   linkedSPL:pharmgxData  ?o .
?s   linkedSPL:supply  ?x3 .
?o   linkedSPL:pharmgxSPLSection  ?o2 .
} order by ?x ?o2 offset  20 limit   10
\end{verbatim}

\end{document}